\documentclass[10pt,aps.prl,twocolumn,superscriptaddress]{revtex4-1} 
\pdfoutput=1
\usepackage{amsmath}
\usepackage{graphicx}
\usepackage{color}
\usepackage{caption}
\usepackage{subcaption}
\usepackage{float}
\usepackage{array}

\begin{document}
\title{Low-Threshold and Narrow Linewidth Diffusive Random Lasing in Rhodamine 6G Dye-Doped Polyurethane with Dispersed ZrO$_2$ Nanoparticles}
\author{Benjamin R. Anderson}
\author{Ray Gunawidjaja}
\author{Hergen Eilers}\email{Corresponding Author: eilers@wsu.edu.}
\affiliation{Applied Sciences Laboratory, Institute of Shock Physics, Washington State University,
Spokane, WA 99210-1695}
\date{\today}

\begin{abstract}
We report on low-threshold and narrow linewidth intensity feedback random lasing in Rhodamine 6G dye-doped polyurethane with dispersed ZrO$_2$ nanoparticles.  Depending on the dye/particle concentration, the lasing threshold is (6.8--15.4) MW/cm$^2$ and the linewidth is (4--6) nm. The lasing threshold as a function of nanoparticle concentration is found to follow a power law with an exponent of $-0.496 \pm 0.010$, which is within uncertainty of Burin {\em et.al.}'s theoretical prediction \cite{Burin01.01}.

\vspace{1em}
{\em OCIS Codes:} (140.2050) Dye lasers; (160.5470) Polymers; (290.4210) Multiple scattering; (160.3380) Laser materials; (290.5850) Scattering, particles.
 
\end{abstract}

\maketitle

\vspace{1em}

\section{Introduction}
Random lasing (RL) is the phenomenon of amplified stimulated emission in a disordered gain medium in which amplification occurs due to scattering within the system rather than due to an optical cavity.  It was first predicted in the late 60's by Letokhov \cite{Letokhov66.01,Letokhov67.01,Letokhov67.02} and finally realized experimentally in the mid 90's \cite{Lawandy94.01}.  These pioneering experiments are characterized by a single narrow (FWHM $\approx 10$ nm) emission peak. This mode of random lasing has been modeled using a light diffusion framework, in which photon propagation is wholly determined by the transport mean free path \cite{Wiersma96.01,Pinheiro06.01,Burin01.01}.  RL in this regime has been labeled as intensity feedback random lasing (IFRL).

Shortly after the initial observation of IFRL a new regime of RL was observed -- known as resonant feedback random lasing (RFRL) -- in which many subnanometer-linewidth peaks emerge on top of the emission spectrum \cite{Ling01.01,Cao03.01,Cao99.01}.  This mode of emission has been found to be inexplicable in the light diffusion framework.  Therefore two different explanations of this phenomenon have been proposed: Anderson localization of light \cite{Cao00.01} and strong scattering resonances \cite{Molen07.01}.   Based on these mechanisms RFRL has been theoretically modeled  using strongly interacting lossy modes \cite{Tureci08.01}, spin-glass modeling of light \cite{Angelani06.01}, condensation of lasing modes \cite{Conti08.01,Leonetti13.03} and Levy-flight scattering \cite{Ignesti13.01}.  Despite these models, it is still difficult to predict the RFRL spectrum as the emission depends on the random distribution of trillions of particles.  This difficulty limits the usefulness of RFRL systems as they cannot 
be 
designed {\em ab initio} to have certain spectral features.  However, recent work by Leonetti and coworkers has shown that while the spectrum can't be determined  {\em ab initio}, it can be controlled using adaptive optic beam shaping.  Using a spatial light modulator (SLM) based pumping scheme Leonetti {\em et al.} were able to successfully control the spectrum and direction of RFRL emission \cite{Leonetti12.02,Leonetti12.03,Leonetti13.02,Leonetti13.01}.

The ability to control the RFRL spectrum is of potential interest in the field of secure authentication, particularly in the implementation of physically unclonable functions (PUFs).  PUFs are physical objects made with large numbers of uncontrollable degrees of freedom (such as scattering materials) which can be used as keys, seals, and markers that are unfeasibly difficult to copy \cite{Goorden13.01,Pappu02.01,Buchanan05.01}. The ability to control the RFRL spectrum can be used to implement PUFs as follows.  First a random lasing PUF is made that can be attached to the object it is to secure.  Once attached the emission spectrum is measured and varied using an SLM to create a challenge (SLM setting) and response (emission spectrum).  Given the random nature of the RL medium, the challenge-response pair will be uniquely related to the PUF.  Any tampering or attempt to copy the PUF would result in the challenge and response decoupling, giving evidence of tampering.

With the end goal of developing RFRL PUFs we investigate RL in Rhodamine 6G dye-doped polyurethane (Rh6G/PU) with dispersed ZrO$_2$ nanoparticles (NPs).  We choose this combination as: (1) Rhodamine 6G is a well known high efficiency laser dye, (2) polyurethane can be formulated as an adhesive (making attachment easy), and (3) ZrO$_2$ has a large refractive index making it a good scattering material.  These properties make  Rh6G+ZrO$_2$/PU an ideal candidate for a real world PUF.  In this current study, we measure the basic RL properties of Rh6G+ZrO$_2$/PU, specifically: the system's lasing regime (IFRL vs RFRL) with no SLM beam shaping and the lasing threshold, line width, and peak location's dependence on dye/particle concentration and pump energy.

\section{Methods}
\subsection{Sample Preparation}
Sample preparation begins by first synthesizing ZrO$_2$ NPs  by forced hydrolysis \cite{Gunawidjaja13.01} with a calcination temperature and time of 600 $^{\circ}$C and 1 h respectively.  SEM measurements of the resulting particles reveal that the particles are spherical with a diameter of 200 nm --300 nm as shown in Figure 1.  

\begin{figure*}
 \centering
 \begin{subfigure}[b]{0.4\textwidth}
 \includegraphics{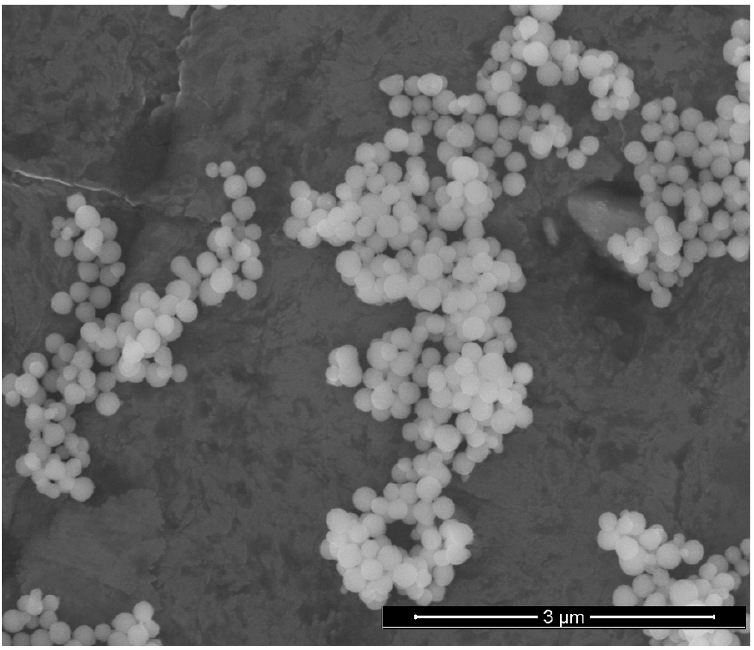}
 \label{fig:nps1}
 \end{subfigure}
 \hspace{4em}
\begin{subfigure}[b]{0.4\textwidth}
 \includegraphics{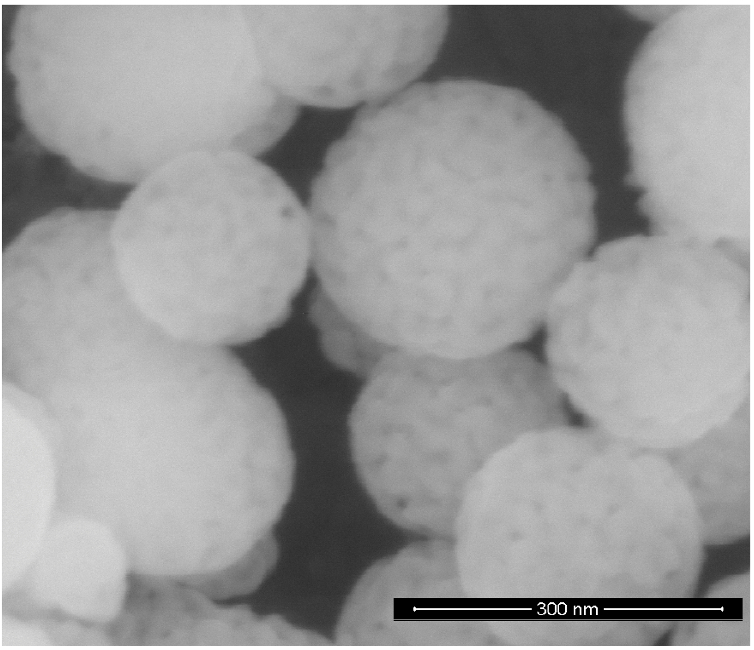}
 \label{fig:nps}
\end{subfigure}
\caption{SEM images of ZrO$_2$ nanoparticles at different magnifications.  The particles are found to be spherical with diameters ranging between 200 nm and 300 nm.}
\end{figure*}

Once the NPs are made, we prepare the dye-doped polymer by first dissolving the appropriate amount of Rhodamine 6G in tetraethylene glycol (TEG). The amounts of Rhodamine 6G added correspond to 0.1 wt\%, 0.5 wt\%, and 1.0 wt\% with respect to the mass of TEG/pHMDI mixture. Next, equimolar amounts of tetraethylene glycol (TEG) and poly(hexamethylene diisocyanate) (pHMDI) are mixed. The TEG (Alfa Aesar, 99\% purity) and pHMDI (Sigma-Aldrich, viscosity is 1,300 cP -- 2,200 cP at 25 $^\circ$C) are both purchased from Sigma-Aldrich.  The viscous mixture is then stirred vigorously and placed in a syringe, which is centrifuged for 3 min at 3000 rpm to remove air bubbles. Once the air bubbles are removed the dye-doped solution is poured into a circular die. Next, we make a dispersion of ZrO$_2$ in 1,4-dioxane and di-n-butyltin dilaurate (DBTDL) with the dioxane at 20 wt\% of the mass of TEG/pHDMI mixture and the DBTDL at 0.1 wt\%. The dispersion is added to the die and the solution is stirred until the mixture 
becomes homogeneous. The mixture is left to cure overnight at room temperature at which point the polyurethane composite is sufficiently robust to be released from the die.

While we set out to make samples with well disperesed nanoparticles, we find that the nanoparticles tend to sink in the polymer solution and aggregate at the bottom of the die. Figure 2 shows optical microscopy images of both the top and bottom of the sample. From these images we see that at the top of the sample there are few visible NP clusters, while the bottom surface is coated in NP clusters. Since RL requires scattering to occur, we use the bottom surface as our incident surface when optically pumping.

\begin{figure*}
 \centering
 \begin{subfigure}[b]{0.4\textwidth}
 \includegraphics{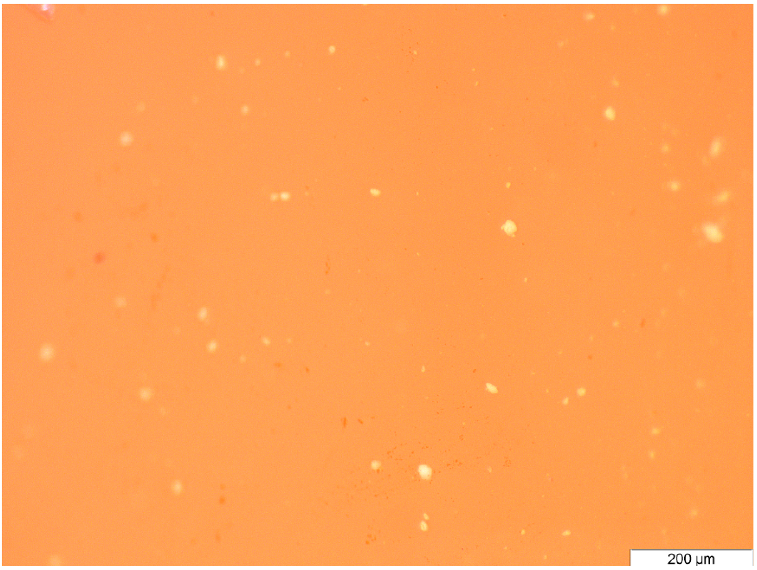}
  \caption[A]{(A) Top}
 \label{fig:top}
 \end{subfigure}
 \hspace{4em}
\begin{subfigure}[b]{0.4\textwidth}
 \includegraphics{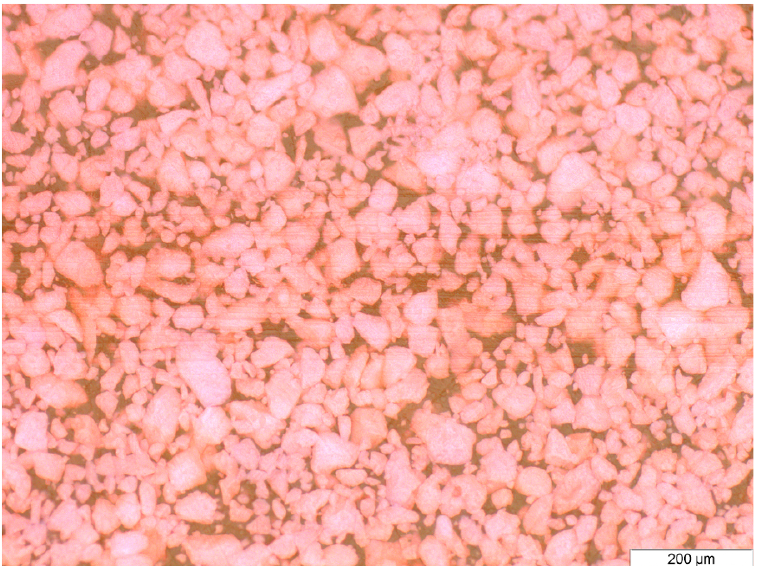}
  \caption[B]{(B) Bottom}
 \label{fig:bottom}
\end{subfigure}
\caption{Optical microscopy images of top and bottom surface of a sample with NP concentration of 2.5 wt\%.  The nanoparticles are found to sink and form aggregates on the bottom of the sample.}
\end{figure*}

\subsection{Experimental Setup}
\begin{figure}
 \centering
 \includegraphics{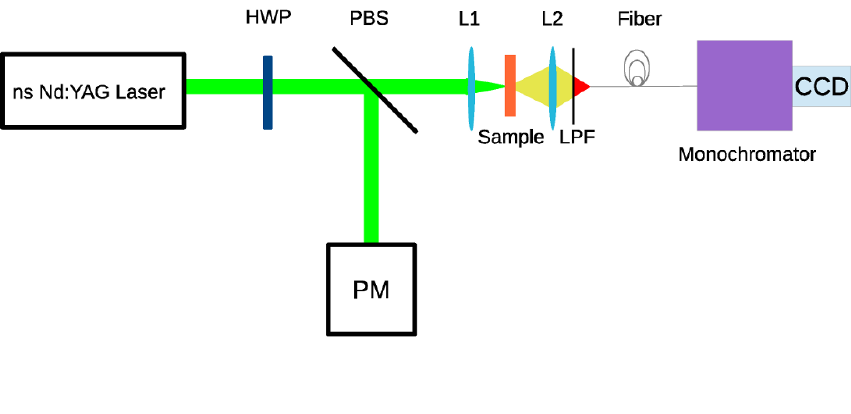}
 \caption{Setup for measuring random lasing. HWP: half-waveplate, PBS: Polarizing beamsplitter, PM: power meter, L1: $f=80$ mm cylindrical lens, L2: $f=35$ mm spherical lens, LPF: longpass filter.}
 \label{fig:setup}
\end{figure}

Figure \ref{fig:setup} shows a schematic of the experimental setup. Emission from a Q-switched frequency doubled Nd:YAG ns laser (Spectraphysics SpectraPro, 532 nm, 10 ns, 10 hz) is passed through a half-waveplate polarizing beamsplitter combination to produce intensity controlled pump light.  The pump light is focused onto the sample using a $f=80$ mm cylindrical lens, producing an elliptical pump spot with a length of 1 cm and width of 1.5 mm.  The emission from the sample is collected using a spherical lens ($f=35$ mm) and coupled into a fiber that is connected to a Princeton Instruments Acton SP-2300i Monochromator with an attached Pixis 2K CCD for spectral measurements.

\section{Results and Discussions}

\begin{figure}
 \centering
 \includegraphics{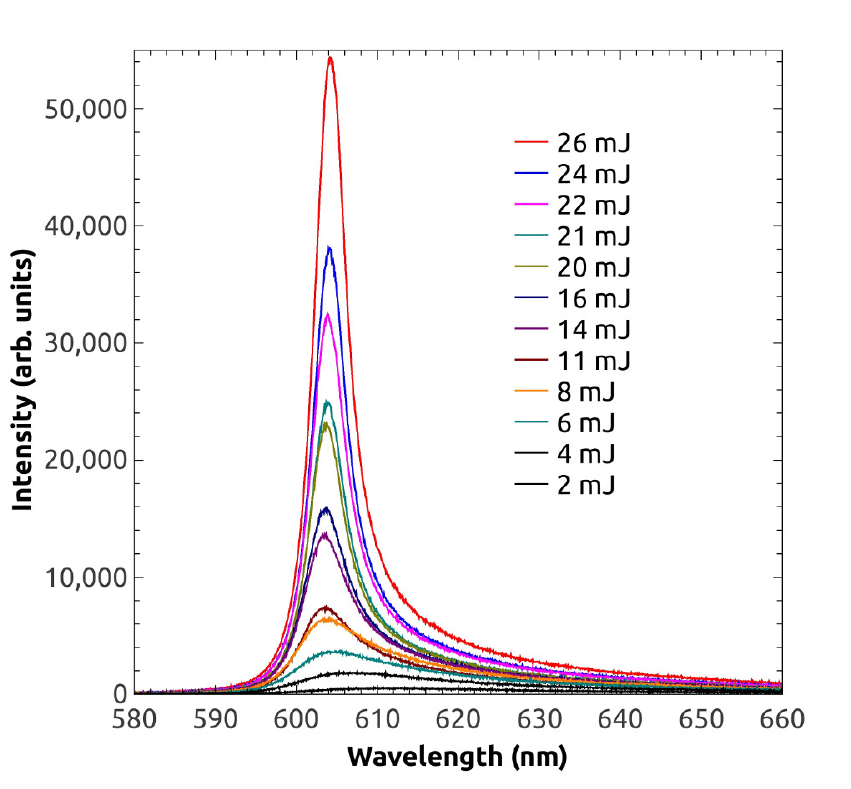}
 \caption{Emission spectra of Rh6G+ZrO$_2$/PU for a dye concentration of 1 wt\% ($2.13\times10^{-2}$M) and nominal particle concentration of 5 wt\% ($1.13 \times 10^{12}$ cm$^{-3}$). }
 \label{fig:spec}
\end{figure}

Emission spectra are measured for each sample at different pump energies, with the spectra being averaged over 100 pump pulses. Figure \ref{fig:spec} shows example spectra at different pump energies for a sample with dye concentration of 1 wt\% ($2.13\times10^{-2}$M) and particle concentration of 5 wt\% ($1.13 \times 10^{12}$cm$^{-3}$).  At low pump energies -- below the lasing threshold -- the emission spectra are found to be broad (FWHM $\approx35$ nm) with the peak wavelength near 615nm.  As the pump energy is increased -- surpassing the lasing threshold -- the peak shifts toward shorter wavelengths ($\sim 600$ nm) and its width narrows drastically (FWHM $\approx 5$nm).  The single-peak narrow linewidth emission from each concentration places the RL emission in the IFRL regime.

\subsection{Linewidth Measurements}

\begin{figure}
 \centering
 \includegraphics{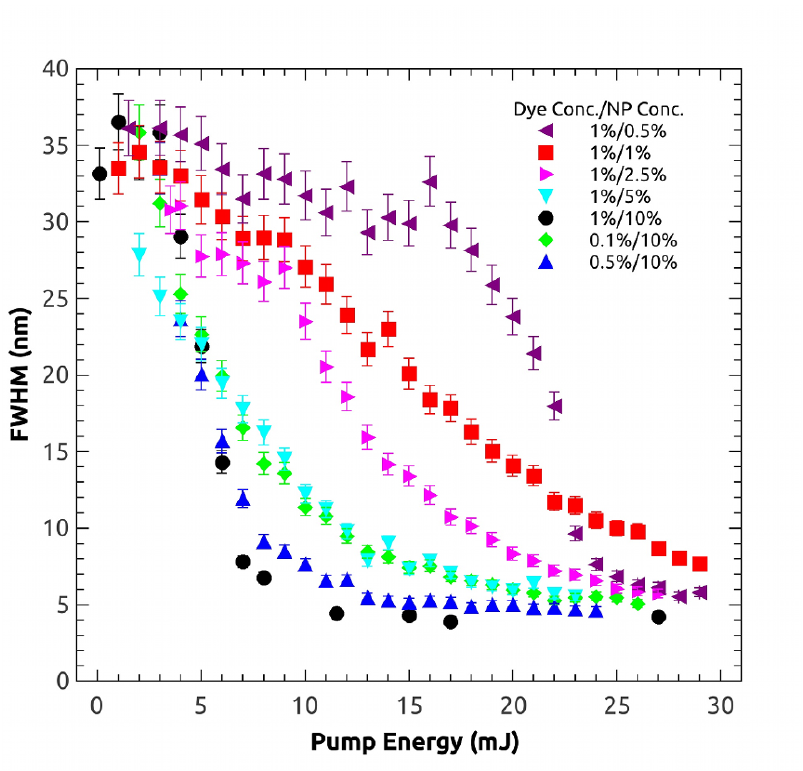}
 \caption{FWHM as a function of pump energy for different dye and NP concentrations.}
 \label{fig:FWHM}
\end{figure}

To determine the linewidth for each concentration and  pump energy, we calculate the FWHM of each emission spectra by direct measurement.  Figure \ref{fig:FWHM} shows the FWHM as a function of pumpenergy for each concentration. The linewidth of each concentration is approximately 35nm for low pump energies and narrows to near 5nm as the pump energy is increased.  Table \ref{Tab:fixeddye} tabulates the FWHM for varying NP concentration at fixed dye concentration and Table \ref{Tab:fixedpart} contains the results for changing dye concentration and constant NP concentration.

Looking at the dye and NP concentration dependence of the curves in Figure \ref{fig:FWHM} it becomes apparent that while both concentrations influence the FWHM, the NP concentration has the greater influence.  Increasing the NP concentration causes the FWHM to collapse to its lasing value at a much faster rate and, once lasing, the FWHM is smaller for higher NP concentrations.  These effects are consistent with previous measurements of other Rh6G NP/host combinations \cite{Dominguez11.01,Chiad11.01,Kitur10.01}.    

The FWHM dependence on NP concentration is to be expected, as light scattering due to the NP gives rise to the non resonant feedback required for RL \cite{Letokhov66.01,Letokhov67.01,Letokhov67.02}. As the NP concentration increases the number of times light is scattered within the sample increases, leading to greater population inversions and stronger lasing.  These results also suggest that the lasing threshold of our system will be strongly dependent on the NP concentration.

\subsection{Lasing Threshold}
\begin{figure}
 \centering
 \includegraphics{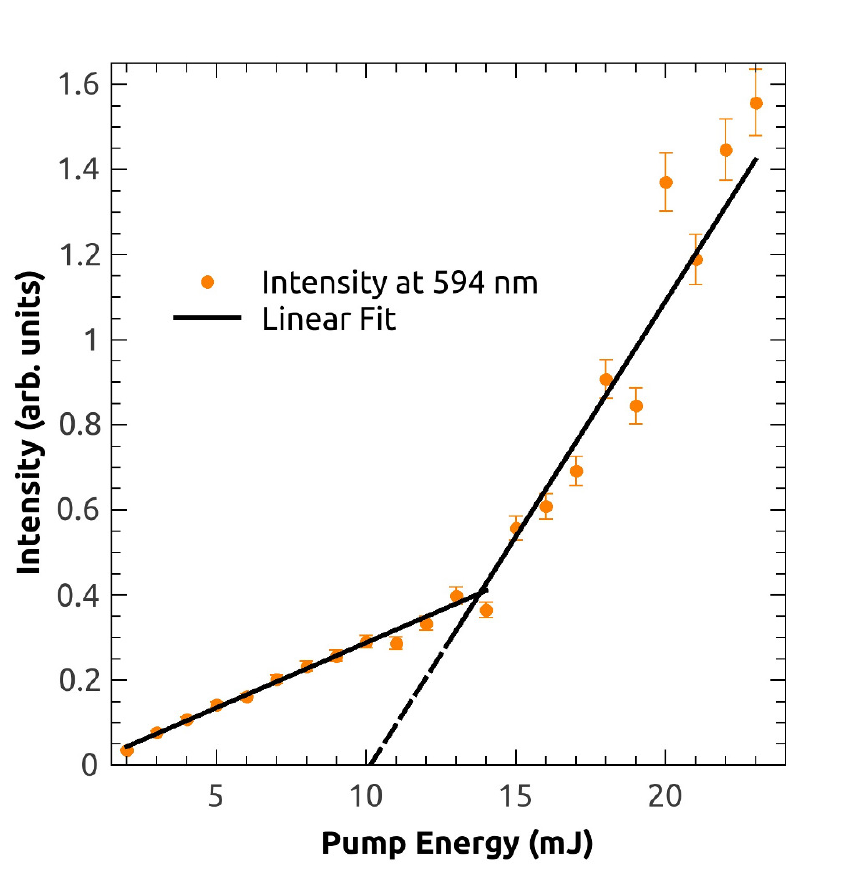}
 \caption{Intensity at lasing peak as a function of pump energy for a sample with dye concentration of 0.1 wt\% ($2.13\times10^{-3}$ M) and nominal NP concentration of 10 wt\% ($2.26 \times 10^{12}$ cm$^{-3}$). Linear fits are performed in the different pump regimes to determine the lasing threshold.}
 \label{fig:intpump}
\end{figure}

To quantify the lasing threshold for each concentration we measure the peak emission intensity as a function of pump energy.  Figure \ref{fig:intpump} shows an example intensity curve as a function of pump energy for a sample with a dye concentration of 0.1 wt\% ($2.13\times10^{-3}$M) and nominal particle concentration of 10 wt\% ($2.26 \times 10^{12}$cm$^{-3}$).  The emission intensity is found to follow a bi-linear function, with the high pump energy emission having a slope approximately ten times larger than the low pump energy emission.  From the emission intensity vs pump energy curve the lasing threshold can be determined by backtracing the high pump energy line fit to where the fit intersects the $x$ axis.  The pump energy location of this intersection is the lasing threshold \cite{Vutha06.01,Kitur10.01}.  Table \ref{Tab:fixeddye} lists the threshold intensities (as well as FWHM) for each particle concentration, while Table \ref{Tab:fixedpart} lists the measured values for each dye concentration.

\begin{table}
\begin{tabular}{|>{\centering\arraybackslash}m{2cm} >{\centering\arraybackslash}m{2.5cm} >{\centering\arraybackslash}m{2.5cm}|}
\hline
 \textbf{ZrO$_2$ Conc. ($10^{12}$cm$^{-3}$)} & \textbf{Threshold Intensity (MW/cm$^2$)} & \textbf{Min. FWHM (nm)}\\ \hline 
 2.26  &   $6.80  \pm 0.65 $ &  4.11 $\pm$  0.34\\ 
  1.13  &   $7.50  \pm 0.94 $  &  4.99 $\pm$ 0.45\\
 0.57  &   $9.44 \pm 0.54 $  &  4.28  $\pm$  0.23\\
  0.23   &   $10.6  \pm 1.2 $  &   6.40  $\pm$  0.26\\ 
 0.11   &   $15.4  \pm  1.1$ &  6.00  $\pm$  0.18 \\ \hline
\end{tabular}
\caption{Random lasing threshold intensity for different concentrations of ZrO$_2$ nanoparticles at fixed dye concentration of 1 wt\% (21.3 $\times$ 10$^{-3}$ M).}
\label{Tab:fixeddye}
\end{table}

\begin{table}
\begin{tabular}{|>{\centering\arraybackslash}m{2cm}  >{\centering\arraybackslash}m{2.5cm} >{\centering\arraybackslash}m{2.5cm}|}
\hline
 \textbf{Dye Conc. (10$^{-3}$M) } & \textbf{Threshold Intensity (MW/cm$^2$)} & \textbf{Min. FWHM (nm)}\\ \hline 
2.13  &  $8.60  \pm  0.56  $  &  5.26  $\pm$  0.53 \\
10.7  &    $7.5 \pm 1.3$ &  4.78  $\pm$  0.15\\
21.3  &    $6.80  \pm 0.65 $ &  4.11 $\pm$  0.34\\  \hline
\end{tabular}
\caption{Random lasing threshold intensity for different concentrations of Rh6G with fixed  nominal ZrO$_2$ concentration of 10 wt\% (2.26 $\times$ $10^{12}$ cm$^{-3}$).}
\label{Tab:fixedpart}
\end{table}

\begin{figure}
 \centering
 \includegraphics{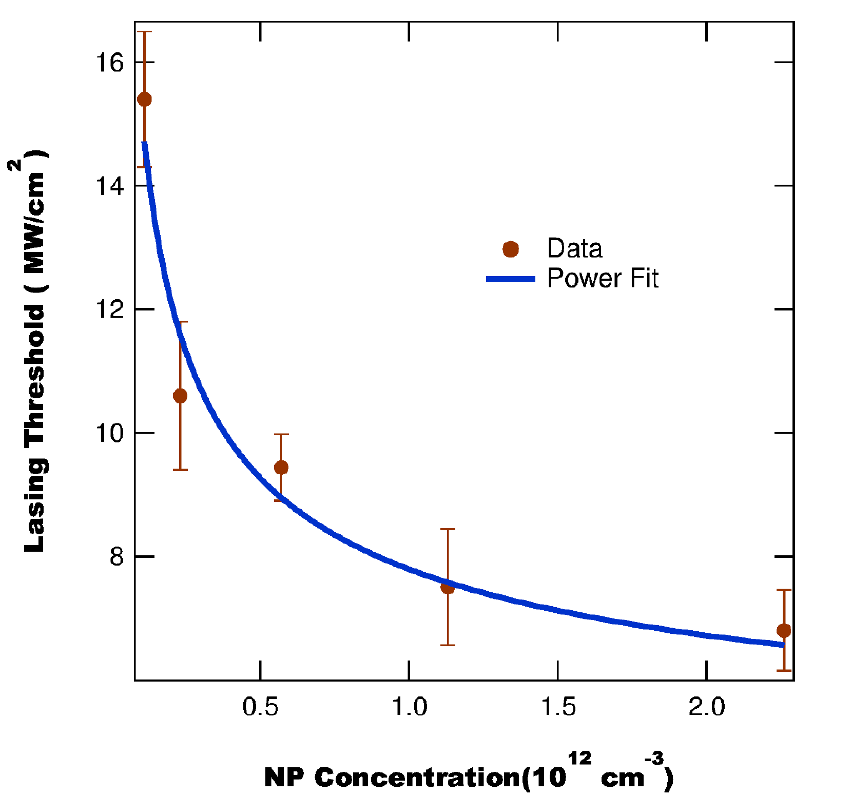}
 \caption{Lasing threshold as a function of ZrO$_2$ nanoparticle concentration.  The lasing threshold decreases as the nanoparticle concentration increases. }
 \label{fig:RLT}
\end{figure}

From Tables \ref{Tab:fixeddye} and \ref{Tab:fixedpart} we see that the lasing threshold decreases as either the dye and/or particle concentration increases.  These results are consistent with our FWHM measurements and previous measurements of RL in different Rh6G dye-doped NP/host combinations \cite{Kitur10.01,Chiad11.01}.  Plotting the threshold values from Table \ref{Tab:fixeddye} as a function of NP concentration (see Figure \ref{fig:RLT}) we find that the threshold depends on the particle concentration via a power function of the form

\begin{equation}
 I_{TH}=Ac^p+I_{0},\label{eqn:fit}
\end{equation}
where $I_{0}$ is the asymptotic threshold, $p$ is the power, and $A$ is the amplitude factor. Using Equation \ref{eqn:fit} to fit the data in Figure \ref{fig:RLT} we find an asymptotic threshold of $I_{TH,0}=4.46  \pm  0.75$ MW/cm$^2$ and an exponential power of $p=-0.496 \pm 0.020$.  

To better understand the meaning of our fit results we consider two different models of IFRL in the light diffusion approximation.  The first by Pinheiro and Sampaio considers a three dimensional RL system, finding  that the random lasing threshold should follow a power law as a function of NP concentration with an exponent of $p=-2/3$ \cite{Pinheiro06.01}.  The second model by Burin considers RL in two dimensions and finds an exponent of $p=-0.5$ \cite{Burin01.01}.  Given that our exponent is within experimental uncertainty of Burin's model, we conclude that our system behaves like a two dimensional system.  We hypothesize that this behavior is due to the NPs aggregating at the incident surface. This aggregation results in gain feedback occuring primarily near the surface, turning the sytem into an effective two-dimensional random laser \cite{Dominguez11.01}.

\subsection{Lasing Peak Location}
\begin{figure}
 \centering
 \includegraphics{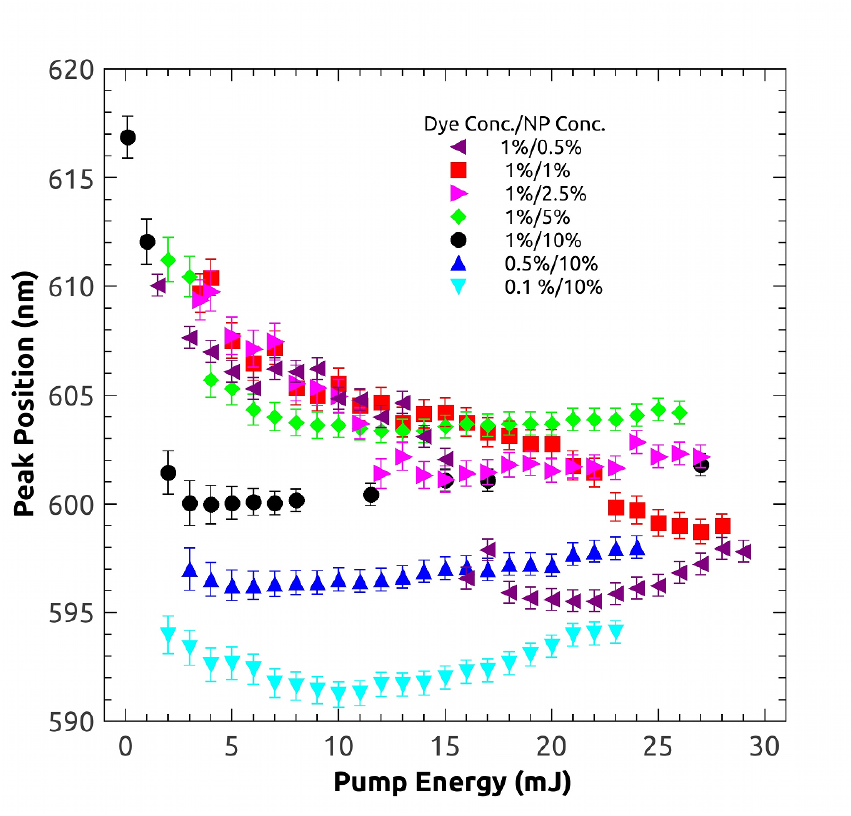}
 \caption{Wavelength of peak emission as a function of pump energy.}
 \label{fig:peak}
\end{figure}

Next we measure the lasing peak wavelength as a function of pump energy and concentration, shown in Figure \ref{fig:peak}. The peak wavelength is found to depend covariantly on the dye concentration, NP concentration and pump energy, with four main features:

\begin{enumerate}
\item There is a large blueshift  as the sample begins to lase.
\item  The emission experiences a redshift as dye concentration increases.
\item Initially, increasing NP concentration causes the peak wavelength to redshift. However, the peak wavelength eventually goes through an inflection point, as shown in Figure  \ref{fig:spconc}.
\item After the initial abrupt blueshift, increasing the pump energy results in a gradual redshift of the lasing peak. 
\end{enumerate}

The initial blueshift is due to the system transitioning from fluorescence into lasing and results from the underlying mechanisms behind each emission.  The other peak wavelength behaviors have been previously observed in other RL systems  \cite{Shuzhen09.01,Vutha06.01}, with different mechanisms proposed for each.  

The redshift with increasing pump energy is hypothesized to be due to photo thermal heating of the sample, as increasing the temperature is known to result in a redshift of the emission spectrum \cite{Vutha06.01}. In the case of the redshift due to increasing dye concentration, the effect occurs due to self-absorption of the emission in the sample \cite{Shuzhen09.01,Ahmed94.01,Shank75.01}.  Since the emission and absorption spectra of Rh6G overlap at shorter wavelengths, the emission in that spectral region is suppressed by self-absorption; while the longer wavelength emission is amplified.  Increasing the dye concentration results in greater absorption and therefore 
greater redshifting.

Self-absorption -- with the addition of scattering effects -- can also be used to explain the inflection point in the peak wavelength as a function of NP concentration.  As the NP concentration increases light is scattered more within the sample leading to a greater pathlength.  Thisw increase results in more self absorption and a red shift of the spectrum.  However, at a certain NP density the scattering will begin to be primarily in the backwards direction, resulting in a decreased penetration depth, leading to less self-absorption and a shift in the spectrum toward shorter wavelengths.

\subsection{Intensity and NP concentration}
Along with self-absorption and scattering affecting the peak wavelength of emission, we also anticipate that these effects will influence the intensity emitted in the forward direction.    To compare the intensity between different NP concentrations we use the spectrally integrated intensity, which makes the analysis insensitive to changes in the peak wavelength.

\begin{figure}
\centering
\includegraphics{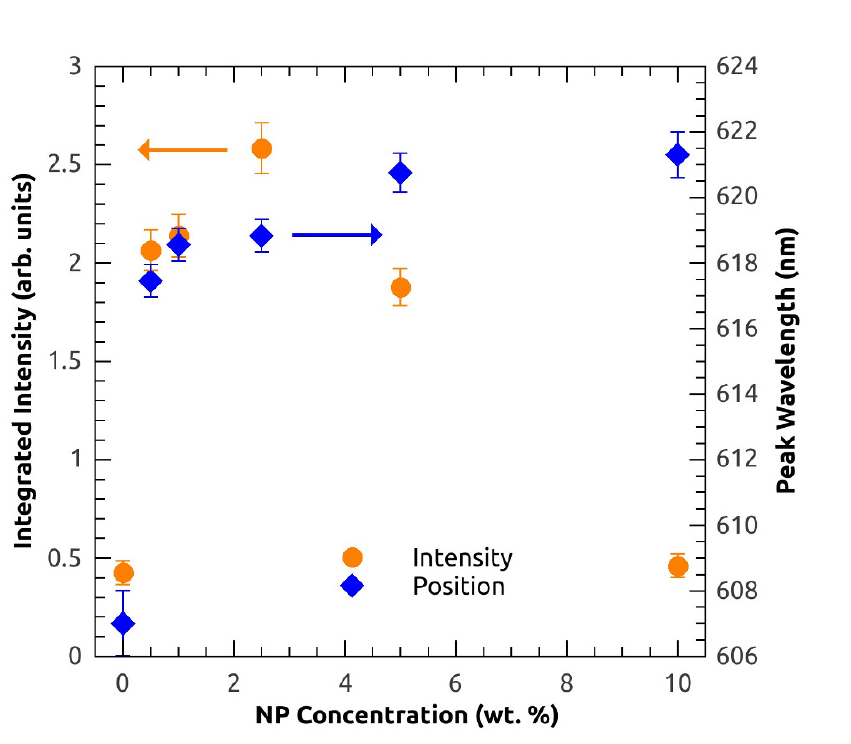}
\caption{Spectrally integrated fluorescence intensity in the forward direction (orange circles) and fluorescence peak position (blue diamonds) as a function of NP concentration.  As the nominal NP concentration increases the peak position is redshifted and the intensity reaches a peak at 2.5 wt\% before declining. }
\label{fig:lpconc}
\end{figure}

\begin{figure}
\centering
\includegraphics{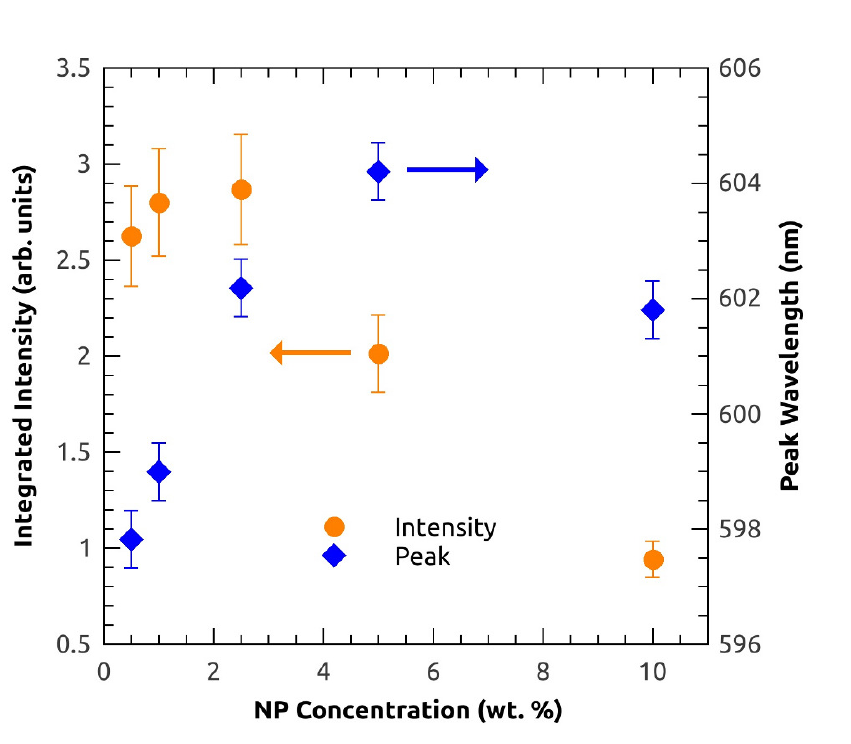}
\caption{Spectrally integrated RL intensity in the forward direction (orange circles) and RL peak position (blue diamonds) as a function of NP concentration.   Zero NP concentration is excluded as it does not display RL.}
\label{fig:spconc}
\end{figure}

To begin we consider low pump energy emission, in which the sample is only fluorescing.  Figure \ref{fig:lpconc} shows the spectrally integrated fluorescence intensity in the forward direction (orange circles) and fluorescence peak position (blue diamonds) as a function of NP concentration.  With the introduction of even a small concentration of NP (0.5 wt\%) both the integrated intensity and peak position change drastically from the dye-doped polymer. The intensity increases by a factor of 4.8  and the peak position experiences an 11 nm redshift.  This effect is due to the inclusion of scattering particles, which increases the pathlength of both pump and emitted light in the sample.  This increase in pathlength causes the pump light to excite more dye molecules and the emitted light to experience more self-absorption, thus leading to an increased emission intensity that is red shifted.  However, as the NP concentration increases further it eventually reaches the point where backscattering begins to 
dominate and the emission intensity in the forward direction decreases as the pump light's penetration depth decreases.  While the effect of transitioning to a new scattering regime affects the fluorescence emission intensity, it does not appear to affect the redshift of the emission, with the peak stabilizing near 621 nm for higher NP concentrations.    This peak wavelength concentration dependence is different than the behavior seen at high pump energies where the peak wavelength goes through an inflection point with increasing concentration as shown in Figure \ref{fig:spconc}. We hypothesize that this difference in the behavior of the peak location is related to the drastically different mechanisms behind fluorescence and random lasing.

While the behavior of the peak wavelength as a function of NP concentration  is different between low and high pump energies, we find that the effect on the integrated intensity as a function of concentration is independent of pump energy.  Figure \ref{fig:spconc} shows the spectrally integrated RL intensity as a function of concentration with the peak intensity occurring at 2.5 wt\%, which is the same behavior measured at low pump energy.  This consistency at low and high pump energies suggests that pathlength variations due to NP concentration are indeed the underlying mechanism behind the changing intensity with NP concentration.

\subsection{Comparisons to Literature}
Rh6G is a widely used dye for random laser studies, with a large number of publications on different combinations of NPs and hosts.  In order to place our host/NP combination in context, we tabulate the minimum reported lasing threshold and FWHM of a representative set of host/NP combinations, listed in Table \ref{tab:comp}.  From Table \ref{tab:comp} we find that the lasing threshold of Rh6G spans several orders of magnitude depending on the host/NP combination, while the FWHM is limited to (4-14) nm.  Additionally, we find that our system has a comparatively low lasing threshold -- with only four combinations having lower thresholds-- and one of the smallest FWHMs.  The extremely small lasing thresholds of the Ag NPs is related to local field enhancements from localized surface plasmon resonances (LSPRs) due to the metallic  NPs \cite{Dice05.01,Dominguez11.01,Meng13.01,Meng09.01,Meng11.01}.


\begin{table*}
 \begin{tabular}{|>{\centering\arraybackslash}m{1.2cm}  >{\centering\arraybackslash}m{2.3cm} >{\centering\arraybackslash}m{2cm} >{\centering\arraybackslash}m{2cm} >{\centering\arraybackslash}m{2.3cm}>{\centering\arraybackslash}m{2.5cm}  >{\centering\arraybackslash}m{1.7cm} >{\centering\arraybackslash}m{1.5cm}>{\centering\arraybackslash}m{1cm}|}
 \hline
  \textbf{NP}  &  \textbf{Host}  &  \textcolor{black}{NP Conc. ($\times 10^{13}$ cm$^{-3}$)}  & \textcolor{black}{Dye Conc. (10$^-2$ M)}  &     \textcolor{black}{\textbf{Index Mismatch} $w$} &\textcolor{black}{\textbf{Particle Diameter} (nm)} &  \textbf{$I_T$ (MW/cm$^2$)}  &  \textbf{FWHM (nm)}  &  \textbf{Ref.}\\ \hline
  ZrO$_2$     &   Polyurethane   & \textcolor{black}{0.226}  &  \textcolor{black}{2.13} &  \textcolor{black}{0.190} &\textcolor{black}{250} & 6.8   &   4.1    &   -  \\
  Latex           &   (poly)MMA-HEMA   &\textcolor{black}{-}  &  \textcolor{black}{0.99} & \textcolor{black}{0.039} &\textcolor{black}{52} &   1200   &   -    &  \cite{Enciso12.01}  \\
  SiO$_2$     &   Cellulose      & \textcolor{black}{-}  &  \textcolor{black}{10$^{-2}$} & \textcolor{black}{0.015} &\textcolor{black}{25} &  2.4    &   5      &   \cite{Santos12.01} \\
  SiO$_2$     &   Silica gel     &  \textcolor{black}{5.3}  &  \textcolor{black}{0.66} &\textcolor{black}{0.030} &\textcolor{black}{3000} &   19.1   &   10     &   \cite{Revilla08.01}\\
      \textcolor{black}{MgF$_2$}  &    \textcolor{black}{Ethanol}  &\textcolor{black}{1.9}  &  \textcolor{black}{0.15} &  \textcolor{black}{0.022}  &    \textcolor{black}{100}  &   \textcolor{black}{27.7}  &    \textcolor{black}{10}  &   \textcolor{black}{\cite{Yi12.01}} \\
       \textcolor{black}{Al$_2$O$_3$}  &    \textcolor{black}{Dimethylsulfide}  &\textcolor{black}{1.9}  &  \textcolor{black}{0.15} &    \textcolor{black}{0.028}  &    \textcolor{black}{100}  &   \textcolor{black}{23.6}  &    \textcolor{black}{10}  &   \textcolor{black}{\cite{Yi12.01}} \\
  TiO$_2$     &   Ethanol/Glycol &\textcolor{black}{-}  &  \textcolor{black}{2} &\textcolor{black}{0.3213} &\textcolor{black}{100} &   114    &   10     &   \cite{Wang98.01} \\
  TiO$_2$     &   Ethanol        &\textcolor{black}{$6.02 \times 10^{4}$}  &  \textcolor{black}{$10^{-3}$} &\textcolor{black}{0.346} &\textcolor{black}{50} &   20     &   9      &   \cite{Chiad11.01}\\
  \textcolor{black}{TiO$_2$}  &    \textcolor{black}{Ethanol}  & \textcolor{black}{1.9}  &  \textcolor{black}{0.15} &   \textcolor{black}{0.346}  &    \textcolor{black}{100}  &   \textcolor{black}{6.0}  &    \textcolor{black}{10}  &   \textcolor{black}{\cite{Yi12.01}} \\
  TiO$_2$     &   PMMA           &\textcolor{black}{$6.02 \times 10^{4}$}  &  \textcolor{black}{$10^{-3}$} &\textcolor{black}{0.3076} &\textcolor{black}{50} &   20     &   14     &   \cite{Chiad11.01}\\
  TiO$_2$     &   Methanol      &\textcolor{black}{3.65}  &  \textcolor{black}{4} &\textcolor{black}{0.356} &\textcolor{black}{250} &   17.4   &   7     &   \cite{Kitur10.01}\\
    \textcolor{black}{Al$_2$O$_3$}  &    \textcolor{black}{Ethanol}  &\textcolor{black}{1.9}  &  \textcolor{black}{0.15} &    \textcolor{black}{0.127}  &    \textcolor{black}{100}  &   \textcolor{black}{18.9}  &    \textcolor{black}{10}  &   \textcolor{black}{\cite{Yi12.01}} \\
  AlN        &   Methanol    &\textcolor{black}{1.9}  &  \textcolor{black}{0.15} &\textcolor{black}{0.236} &\textcolor{black}{100} &   11       &    10   &   \cite{Yi12.01} \\
  AlN        &   Ethanol       &\textcolor{black}{1.9}  &  \textcolor{black}{0.15} &\textcolor{black}{0.225} &\textcolor{black}{100} &   13.3      &    11   &   \cite{Yi12.01} \\
  AlN        &   Glycol      &\textcolor{black}{1.9}  &  \textcolor{black}{0.15} &\textcolor{black}{0.202} &\textcolor{black}{100} &   15.7       &    10   &   \cite{Yi12.01} \\
  AlN        &   Dimethylsulfide&\textcolor{black}{1.9}  &  \textcolor{black}{0.15} &\textcolor{black}{0.188} &\textcolor{black}{100} &   17.86       &    10   &   \cite{Yi12.01} \\
  Ag          &   Methanol       &\textcolor{black}{2.3}  &  \textcolor{black}{0.1} &\textcolor{black}{0.824} &\textcolor{black}{55} &   0.3    &   5      &   \cite{Dice05.01}\\
  Ag          &   PMMA           &\textcolor{black}{1.3}  &  \textcolor{black}{1} &\textcolor{black}{0.841} &\textcolor{black}{12} &   0.05   &   10     &   \cite{Dominguez11.01}\\ 
  Ag          &   PMMA fiber   &\textcolor{black}{542}  &  \textcolor{black}{$4 \times 10^{-5}$} &\textcolor{black}{0.841} &\textcolor{black}{5} &  471  &  4   &   \cite{Sebastian14.01} \\ \hline 
 \end{tabular}
\caption{Comparison of reported lasing thresholds and FWHMs for Rhodamine 6G in different hosts and with different nanoparticles.  For this comparison we only consider IFRL results as RFRL results in sub-nanometer line widths.}
\label{tab:comp}
\end{table*}

\textcolor{black}{Possible explanations for the large variations in the measured lasing thresholds across studies include different dye concentrations \cite{Vutha06.01,Ahmed94.01,Shank75.01} and a wide variation in scattering lengths of the various materials.  The three main factors affecting the scattering length are the NP size, NP concentration, and the NP/host refractive index mismatch, $w$, which is given by, }
\begin{align}
\textcolor{black}{w=\frac{|n_1-n_0|}{n_1+n_0},}
\end{align}
\textcolor{black}{where $n_1$ is the index of refraction of the NPs and $n_0$  is the refractive index of the host.}

\textcolor{black}{Previously, Yi and coworkers measured the effect of differing index mismatch on the random lasing threshold by measuring samples using different NPs and hosts \cite{Yi12.01}.  They found that the lasing threshold is inversely related to the index mismatch, with small $w$ samples having larger lasing thresholds than samples with large mismatches. While this trend can be seen for some of the samples in Table \ref{tab:comp}, there are many exceptions to the trend, with the most striking examples found by considering the TiO$_2$ NP samples.  TiO$_2$ is has one of the largest refractive indices in nature and thus results in some of the largest index mismatches.  However, the majority of the TiO$_2$ NP samples have larger lasing thresholds than other materials (ZrO$_2$ and AlN NPs for example) which have smaller index mismatches.}

\textcolor{black}{In several cases the high lasing thresholds of the TiO$_2$ NP samples can be attributed to low dye concentrations, while in others it appears that small NPs result in higher thresholds.  This can be understood as the scattering length is inversely related to the particle size in Mie scattering theory \cite{Mie08.01}, so smaller NPs will result in longer scattering lengths and therefore higher thresholds.  Additionally, the NP size appears to have an effect on the lasing FWHM, as the two systems containing particles of a diameter of 250 $\mu$m have the smallest FWHMs (our study and \cite{Kitur10.01}).}

\textcolor{black}{While scattering length and dye concentration comparisons explain most of the variations in literature, there are still some anomalies that still need to be explained, such as SiO$_2$/Celluose's \cite{Santos12.01}  and ZrO$_2$/PU's low thresholds and narrow linewidths.  Given limited information in Santos {\em et al.}'s paper, we cannot comment on possible mechanisms behind their results.  However, for ZrO$_2$/PU, we hypothesize that the accumulation of the ZrO$_2$ NPs at the surface of the sample (see Figure \ref{fig:bottom}) results in a drastically smaller scattering length than the nominal concentration would suggest.  From Mie scattering theory \cite{Mie08.01} and the nominal concentration, the scattering length should be on the order of several microns.  However,  microscopic images of the sample surface give an estimated inter-NP distance on the order of 10's of nm, which presents strong evidence for a much shorter scattering length than estimated from the nominal concentration.  In order to test this hypothesis we are preparing to perform surface scattering length measurements, first demonstrated by Leonetti and coworkers\cite{Leonetti11.01}, as well as producing more uniformly disperse samples for testing.  If the uniformly disperse samples result in drastically different lasing thresholds and FWHM, it would suggest that our current results arise due to the surface agglomeration of the NPs.}

\section{Conclusions}
The longterm goal of our study is to develop RFRL PUFs using dye/NP-doped polymers.  As a starting point we developed Rh6G+ZrO$_2$/PU samples with varying dye and NP concentrations.  For the concentrations tested, Rh6G+ZrO$_2$/PU is found to exhibit RL in the IFRL regime, with the emission properties (threshold, FWHM, peak location, spectrally integrated intensity) dependent on both dye and particle concentration.  The lasing threshold and FWHM are found to decrease as the NP concentration increases, with the lasing threshold found to follow a power function with an exponent of $p=-0.496 \pm 0.020$. This exponent is within uncertainty of the prediction for two-dimensional RL which is due to the NPs aggregating at the bottom surface of our samples during preparation.  While the lasing threshold and FWHM are found to be monotonic with NP concentration, the peak location and spectrally integrated intensity are both found to go through inflection points as the NP concentration is increased.  The proposed 
mechanism for both inflection points is related to how the NP concentration changes the total pathlength in the sample.  

We also find that, over the range of concentrations tested, the lasing threshold is (6.8-15.4) MW/cm$^2$ and the FWHM is (4-6) nm.  These results place the ZrO$_2$/PU combination among the lowest threshold NP/host combinations, as well as one of the smallest linewidth combinations.  

Finally, while the current sample formulation and pump scheme only produced IFRL, the lasing characteristics of Rh6G+ZrO$_2$/PU are encouraging for future development of this system into RFRL PUFs.  We are currently working on testing higher concentration samples, different metallic based NPs, a wide range of NP sizes,  and spatial light modulator controlled pumping in order to try and produce controlled RFRL.  Additionally, we are exploring how different formulations of polyurethane affect the RL properties.

\acknowledgements
This work was supported by the Defense Threat Reduction Agency, Award \# HDTRA1-13-1-0050 to Washington State University.

\bibliographystyle{osajnl}

\end{document}